\documentclass[final,5p,times,twocolumn,authoryear]{elsarticle}

\usepackage{amssymb}
\usepackage{lipsum}
\usepackage{url}
\usepackage{hyperref}
\usepackage{xcolor}
\usepackage{amsmath}
\usepackage{aas_macros}
\usepackage{diagbox}

\newcommand{\tim}{\textcolor{black}}
\newcommand{\revx}{\textcolor{black}}
\newcommand{\rev}{\textcolor{black}}

\journal{Astronomy $\&$ Computing}

\begin{document}

\begin{frontmatter}

\title{Real-bogus scores for active anomaly detection}

\author[sai,phfmsu]{T.~A.~Semenikhin\corref{cor}}
\cortext[cor]{Corresponding author.}
\ead{semenikhintimofey@gmail.com}
\author[sai,hse]{M.~V.~Kornilov}
\author[clermont,sai]{M.~V.~Pruzhinskaya}
\author[sai]{A.~D.~Lavrukhina}
\author[clermont]{E.~Russeil}
\author[clermont]{E.~Gangler}
\author[clermont]{E.~E.~O.~Ishida}
\author[ind]{V.~S.~Korolev}
\author[mcwilliams,urbana]{K.~L.~Malanchev}
\author[iki,sai]{A.~A.~Volnova}
\author[surrey]{S.~Sreejith}
\author[]{(The SNAD team)}

\affiliation[sai]{
            organization={Lomonosov Moscow State University, Sternberg Astronomical Institute},
            addressline={Universitetsky pr.~13}, 
            city={Moscow},
            postcode={119234}, 
            country={Russia}}

\affiliation[phfmsu]{
            organization={Lomonosov Moscow State University, Faculty of Physics},
            addressline={Leninskie Gory~1-2},
            city={Moscow},
            postcode={119991},
            country={Russia}
}

\affiliation[hse]{
            organization={National Research University Higher School of Economics},
            addressline={21/4 Staraya Basmannaya Ulitsa}, 
            city={Moscow},
            postcode={105066}, 
            country={Russia}}

\affiliation[clermont]{
            organization={Université Clermont Auvergne, CNRS/IN2P3, LPCA},
            city={Clermont-Ferrand},
            postcode={63000}, 
            country={France}}

\affiliation[ind]{
            organization={Independent researcher},
            }

\affiliation[mcwilliams]{
            organization={McWilliams Center for Cosmology and Astrophysics, Department of Physics, Carnegie Mellon University},
            city={Pittsburgh},
            postcode={PA 15213}, 
            country={USA}}

\affiliation[urbana]{
            organization={Department of Astronomy, University of Illinois at Urbana-Champaign},
            addressline={1002 West Green Street}, 
            city={Urbana},
            postcode={IL 61801}, 
            country={USA}}

\affiliation[iki]{
            organization={Space Research Institute of the Russian Academy of Sciences (IKI)},
            addressline={84/32 Profsoyuznaya Street}, 
            city={Moscow},
            postcode={117997}, 
            country={Russia}}

\affiliation[surrey]{
            organization={Physics Department, University of Surrey},
            addressline={Stag Hill Campus, GU2 7XH}, 
            city={Guildford},
            country={UK}}

\begin{abstract}
In the task of anomaly detection in modern time-domain photometric surveys, the primary goal is to identify astrophysically interesting, rare, and unusual objects among a large volume of data. Unfortunately, artifacts --- such as plane or satellite tracks, bad columns on CCDs, and ghosts --- often constitute significant contaminants in results from anomaly detection analysis. In such contexts, the Active Anomaly Discovery (AAD) algorithm allows tailoring the output of anomaly detection pipelines according to what the expert judges to be scientifically interesting. We demonstrate how the introduction real-bogus scores,  obtained from a machine learning classifier, improves the results from  AAD. Using labeled data from the SNAD ZTF knowledge database, we train four real-bogus classifiers: XGBoost, CatBoost, Random Forest, and Extremely Randomized Trees. All the models perform real-bogus classification with similar effectiveness, achieving ROC-AUC scores ranging from 0.93 to 0.95. Consequently, we select the Random Forest model as the main model due to its simplicity and interpretability. The Random Forest classifier is applied to 67 million light curves from ZTF DR17. The output real-bogus score is used as an additional feature for two anomaly detection algorithms: static Isolation Forest and AAD. 
\rev{The number of artifacts detected by both algorithms decreases significantly with the inclusion of the real-bogus score in cases where the feature space regions are densely populated with artifacts. However, it remains almost unchanged in scenarios where the overall number of artifacts in the outputs is already small.}
We conclude that incorporating the real-bogus classifier result as an additional feature in the active anomaly detection pipeline reduces the number of artifacts in the outputs, thereby increasing the incidence of  astrophysically interesting objects presented to human experts.

\end{abstract}


\end{frontmatter}

\section{Introduction}
\label{sec:introduction}
Modern sky surveys offer the opportunity to automatically observe a vast number of astrophysical objects across different classes. For instance, the Zwicky Transient Facility (ZTF;~\citealt{2019PASP..131a8002B}) generates about 2 terabytes of data per night, and the upcoming Vera C. Rubin Observatory Legacy Survey of Space and Time (LSST;~\citealt{2009arXiv0912.0201L}) is expected to increase this amount by an order of magnitude. When dealing with such large volumes of data, machine learning (ML) methods are unavoidable, especially in tasks such as classification and anomaly detection (e.~g.,~\citealt{2019arXiv190407248B,2019NatAs...3..680I,2020ApJS..249...18C,2022MNRAS.513.5505M,2023A&A...672A.111P};~\tim{\citealt{electronics11193105,electronics11244210}}).
However, the task of anomaly detection is particularly challenging due to the inherent nature of searching for statistical deviations in the data, which often results in a significant number of outliers being artifacts of automatic image processing, such as defocusing, bad columns, bright star spikes, ghosts, etc. For example, in \cite{2021MNRAS.502.5147M}, 68 percent of outliers found by anomaly detection algorithms in the third ZTF data release (DR) were identified as bogus light curves. Around 80 instances of data artifacts were discovered as unwanted contaminants during the search for gravitational self-lensing binaries with ZTF~\citep{2023arXiv231117862C}. Similarly, \cite{2021AJ....162..206S} encountered an overwhelming amount of bogus among the 8809 anomalies selected while searching for Changing-state \tim{active galactic nuclei} in the ZTF DR5.

Given limited human resources, it is crucial to avoid spending time on non-astrophysical candidates. Active anomaly detection algorithms partly address this issue by incorporating feedback from human experts, which helps fine-tuning the algorithm and reduces the number of astrophysically uninteresting objects~\citep{2021A&A...650A.195I}. However, artifacts, due to their diversity, can occupy different regions of the parameter space, and even active algorithms cannot guarantee their elimination from outliers after several iterations.

One way to approach this problem is to make a real-bogus classification and then reject all objects that the classifier considered as artifacts. For example, this was done for Nearby Supernova Factory~\citep{2002SPIE.4836...61A} by~\cite{2007ApJ...665.1246B} and other time domain surveys such as The Palomar Transient Factory (PTF;~\citealt{2013MNRAS.435.1047B}), the Dark Energy Survey (DES;~\citealt{2015AJ....150...82G}), the Panoramic Survey Telescope and Rapid Response System (Pan-STARRS;~\citealt{2015MNRAS.449..451W}). Nowadays, neural network approaches are actively used for this purpose. For example,~\cite{10.1093/rasti/rzae027} used difference imaging and convolutional neural networks for real-bogus classification in the Asteroid Terrestrial-impact Last Alert System (ATLAS;~\citealt{2018PASP..130f4505T}).~\cite{Acero-Cuellar_2023} used data from DES and investigated the performance of convolutional neural networks without using template subtraction. Deep learning methods are also used on ZTF survey data in real-bogus classification tasks (e.~g.,~\citealt{2019MNRAS.489.3582D,2021AJ....162..231C,rb_nn}).

The SNAD team\footnote{\url{https://snad.space/}}  has been working on the development and adaptation of anomaly detection techniques for astronomical data since 2018. 
In this work, we present and integrate into the SNAD pipeline a novel approach designed to reduce even further the number of artifacts presented to the expert in sequential iterations of the adaptive learning cycle. The idea involves the development of a real-bogus classifier whose predictions are used as an additional feature for each data instance. These enhanced data are then fed into the Active Anomaly Discovery (AAD;~\citealt{Das:16,2021A&A...650A.195I}) algorithm. It has been demonstrated that with this approach, AAD learns significantly faster to avoid returning artifacts to the expert, thus saving the expert's time.

The article is structured as follows: Section~\ref{sec:data} describes the data used for training and testing. Section~\ref{sec:rb_clf} is dedicated to the construction of real-bogus classifiers. Section~\ref{sec:anomaly_detection} presents a comparison of the performance of anomaly detection methods with and without real-bogus scores. Finally, Sections~\ref{sec:discussion} and~\ref{sec:conclusions} contain the discussion and conclusions, respectively.

\section{Data}
\label{sec:data}

The classifier is developed to \tim{improve the results} of anomaly detection in the data from \tim{ZTF}\footnote{\url{https://www.ztf.caltech.edu/}}, an automated wide-field sky survey. ZTF is conducted at the Palomar Observatory in California, USA, using a 1.26-m Samuel Oschin telescope with a field of view of approximately $47$ square degrees. The survey operates in three photometric passbands ($zg$, $zr$, $zi$,~\tim{\citealt{2019PASP..131a8002B}}). Data from ZTF are provided in two formats: alerts and data releases. Alerts are issued in real-time, while data releases contain light curves of variable objects over the entire observation period. In this work, we use the $zr$-band light curves from the data releases only.

\subsection{Real-bogus dataset}
\begin{figure}
	\centering 
	\includegraphics[width=0.45\textwidth]{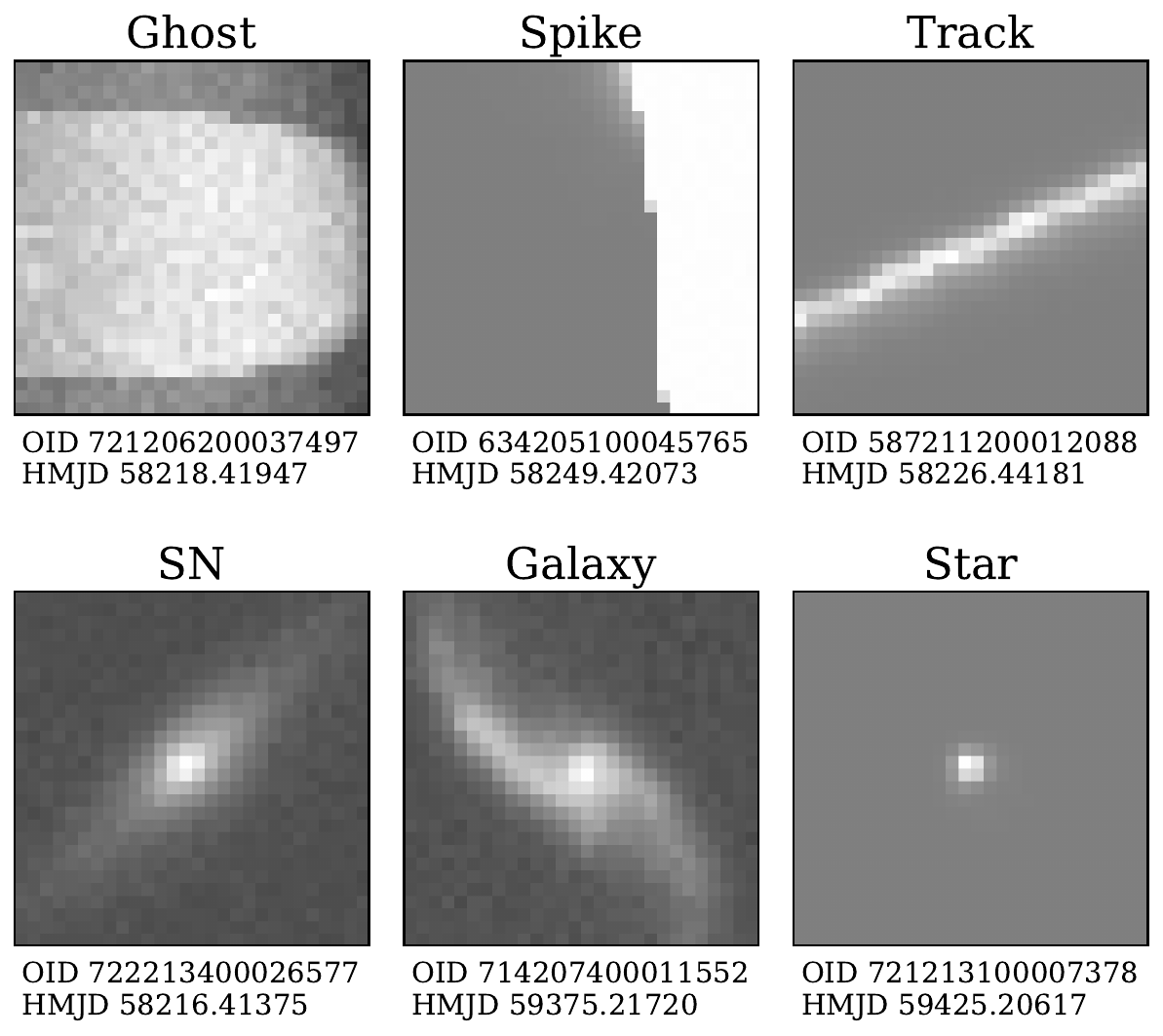}	
	\caption{Object examples of different classes from ZTF data. Each image has size of 28~px $\times$ 28 px, $1~{\rm px} = 1.012''$.} 
	\label{fig:artfs}
\end{figure}

The anomaly knowledge base (AKB; see Section 4 in~\citealt{2023PASP..135b4503M}) \revx{was constructed thanks to the efforts of multiple experts throughout the last 6 years of activities of the } SNAD team. \revx{Their evaluation of each object was recorded} during the scrutiny of the outputs of anomaly detection algorithms runs on various ZTF DRs. \revx{As a consequence, our data base is a unique source labels populating sparse regions of the initial parameter space \citep[e.g., previous SNAD results include ][]{2021A&A...650A.195I,2021MNRAS.502.5147M,2022NewA...9601846A,2023A&A...672A.111P,2024MNRAS.533.4309V}}. Each ZTF object, identified by the object ID (OID), is classified according to the SNAD classification scheme (see examples in Fig.~\ref{fig:artfs}) based on extensive analysis, including visual screening, literature review, photometric model fitting, and follow-up observations. As of July 2023, the database comprised 3311 objects: 1646 artifacts and 1665 astrophysical objects of different types. We use this dataset, referred to as the \textit{real-bogus dataset}, to train and test real-bogus classifiers.

\subsection{Target dataset}
For the anomaly detection task we used light curves from ZTF DR17 within the MJD interval $ \left[59184; 59524 \right]$ subjected to the following cuts: at least 100 observations and $\left| b \right|$ $> 15^{\circ}$, where $b$ is the galactic latitude. This resulted in approximately 67 millions OIDs  that make up the \textit{target dataset}. We ran the Isolation Forest and the Active Anomaly Discovery algorithm on this dataset.

\subsection{Light curve features}\label{subsec:lc}
For each light curve we extracted 54 features using the {\tt lightcurve} package described at~\cite{2021MNRAS.502.5147M}. Some of them have a simple form, such as flux amplitude or mean value. Others have a more complex dependency with the light curve shape, such as skewness or optimized parameters of the Bazin function from~\cite{bazin}. We name this feature set \textit{initial}.

\section{Real-bogus classifiers}
\label{sec:rb_clf}
As classifiers, we consider 4 of the most popular models for working with tabular data. Two of them are based on gradient boosting~\citep{friedman2001greedy}: XGBoost~\citep{Chen_2016} and CatBoost~\citep{dorogush2018catboost}. The others have simpler architectures and interpretability: Random Forest~\citep{rand} and Extremely Randomized Trees (ExtraTrees,~\citealt{extratrees}). The models' parameters were tuned using {\tt Optuna}\footnote{\url{https://optuna.org/}} framework (see Table~\ref{tab:params}). When selecting parameters, the algorithm aimed to maximize accuracy. We choose this metric so it does not coincide with our main validation metric (Receiver Operating Characteristic -- Area Under the Curve (ROC-AUC)). 
\begin{table}
{
\centering
\begin{tabular}{l r} 
    \hline
   \multicolumn{2}{c}{XGBoost} \\
    \hline
    {\tt booster} & {\tt dart}\\
    {\tt lambda} & 0\\
    {\tt alpha} & 0\\
    {\tt subsample} & 0.79\\
    {\tt colsample\_bytree} & 0.66\\
    {\tt max\_depth} & 9\\
    {\tt min\_child\_weight} & 3\\
    {\tt eta} & 0.02\\
    {\tt gamma} & 0\\
    {\tt grow\_policy} & {\tt lossguide}\\
    {\tt sample\_type} & {\tt uniform}\\
    {\tt normalize\_type} & {\tt tree}\\
    {\tt rate\_drop} & 0.1\\
    {\tt skip\_drop} & 0 \\
    \hline
    \multicolumn{2}{c}{CatBoost} \\
    \hline
    {\tt objective} & {\tt CrossEntropy}\\
    {\tt colsample\_bylevel} & 0.07\\
    {\tt depth} & 7\\
    {\tt boosting\_type} & {\tt Plain}\\
    {\tt bootstrap\_type} & {\tt Bayesian}\\
    {\tt bagging\_temperature} & 1.53 \\
    \hline
    \multicolumn{2}{c}{Random Forest} \\
    \hline
    {\tt max\_depth} & 18 \\
     {\tt n\_estimators} & 830 \\
     \hline
    \multicolumn{2}{c}{Extremely Randomized Trees} \\
    \hline
    {\tt max\_depth} & 39\\
    {\tt n\_estimators} & 251\\
    {\tt max\_features}$^\dagger$ & 1 \\
    \hline
\end{tabular}\par
}
\caption{Parameters for real-bogus classifiers optimized with {\tt Optuna}. $^\dagger$This parameter was not varied and set equal to 1.}
\label{tab:params}
\end{table}

We use objects and corresponding labels (0 stands for an artefact, 1 -- astrophysical object) from the real-bogus dataset for training classifiers. The real-bogus classifier works in the following way: we feed light curve features to one of the classical machine learning method listed above, which returns a real-bogus score in the range $[0, 1]$. The closer the number is to 0, the more confident the  model is that the light curve represents an artefact.

Since the real-bogus dataset is relatively small (3311 objects), we use a k-fold ($ k = 5 $) cross-validation  when training the classifiers. In this method, we divide the available data into $k$ subsets. Then, a model is trained on union of $k-1$ subsets and evaluated on the remaining one, called the test set. This process is repeated $k$ times, each time using a different subset as the test set. As a result, $k$ model quality metric estimates are obtained, which are then averaged to compute the final estimate. This approach allows us to avoid overfitting  and to obtain a more reliable estimate\tim{s}.

\subsection{Validation of classifiers}
\begin{figure}
	\centering 
	\includegraphics[width=0.45\textwidth]{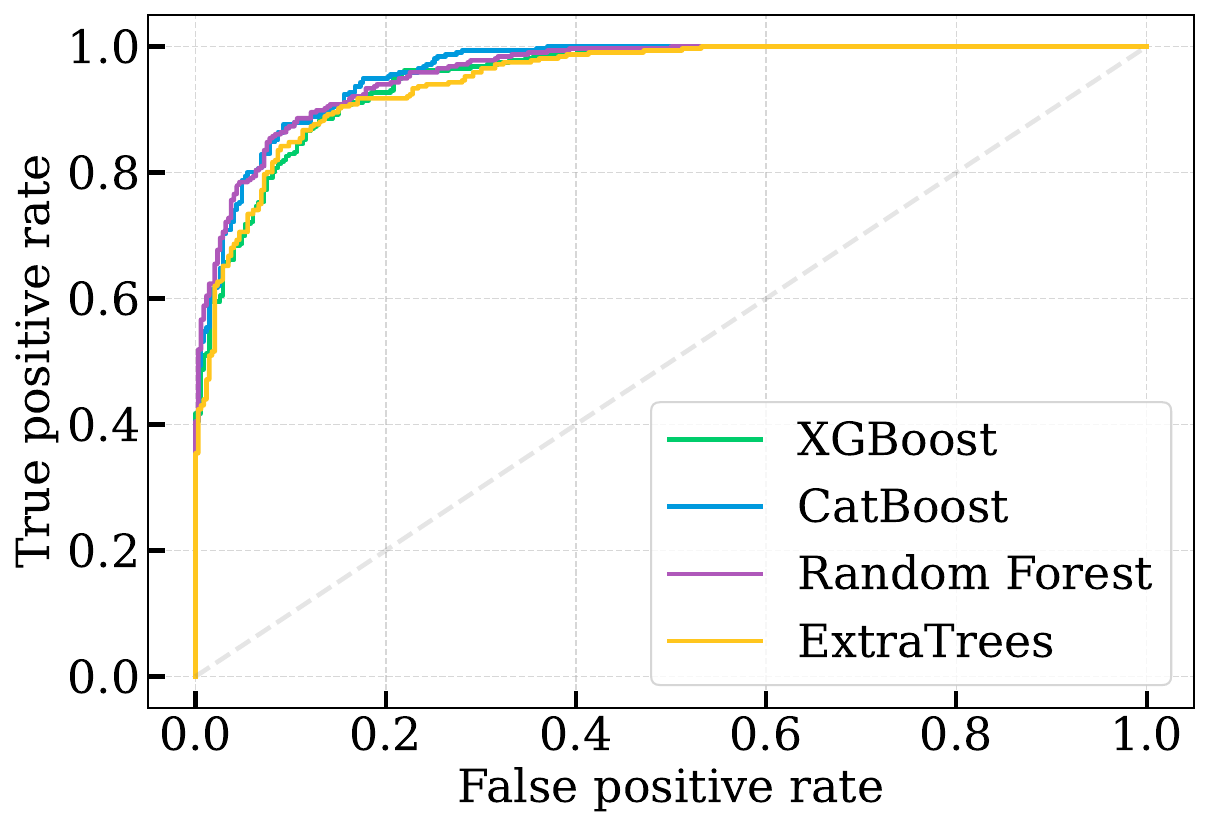}	
	\caption{ROC-curve examples of trained real-bogus classifiers for one of the cross-validation splits. \tim{The grayed diagonal dash line corresponds to a random binary classifier. The closer the ROC curve is to the diagonal line, the worse this classifier is.}
} 
	\label{fig:roc_curves}
\end{figure}

In order to determine the predicted class of an object, it is necessary to choose a threshold for the real-bogus score. However, in this work, the prediction of the classifier serves as an additional source of information for the Active Anomaly Discovery algorithm (see Section~\ref{sec:anomaly_detection}). In this case, the determination of a specific threshold is the prerogative of the AAD algorithm based on expert feedback. Moreover, technically, such a threshold may vary for different regions of the feature space. Therefore, we chose ROC-AUC as a main quality metric for the real-bogus classifiers. 
As additional quality metrics, we also considered Accuracy and F1-score with the standard threshold set to 0.5 as well as ROC-curves (see Fig.~\ref{fig:roc_curves}).  

From Table~\ref{tab:metrics}, which presents the final values of the quality metrics  averaged over the test sets, and Fig.~\ref{fig:roc_curves}, it can be seen that the 4 considered models achieve similar performances. Therefore, we select the Random Forest as the main model because it is the simplest and most interpretable algorithm. All the real-bogus scores presented below were obtained using this model.
\begin{table}
\begin{tabular}{l c c r} 
    \hline
    Model name & ROC-AUC & Accuracy & F1-score \\
    \hline
    Random Forest & 0.94 $\pm$ 0.01 & 0.87 $\pm$ 0.02 & 0.87 $\pm$ 0.02\\
    ExtraTrees & 0.93 $\pm$ 0.01 & 0.85 $\pm$ 0.01 & 0.85 $\pm$ 0.01\\
    XGBoost & 0.93 $\pm$ 0.01 & 0.85 $\pm$ 0.02 &  0.85 $\pm$ 0.01\\
    CatBoost & 0.95 $\pm$ 0.01 & 0.87 $\pm$ 0.02 & 0.87 $\pm$ 0.02 \\
    \hline
\end{tabular}
\caption{Real-bogus classifiers results. The metric values are averaged over 5 test sets, as well as the standard deviation.
}
\label{tab:metrics}
\end{table}

\subsection{Inference}
\begin{figure*}
\centering
\begin{minipage}{0.495\linewidth}
\includegraphics[width=1\linewidth]{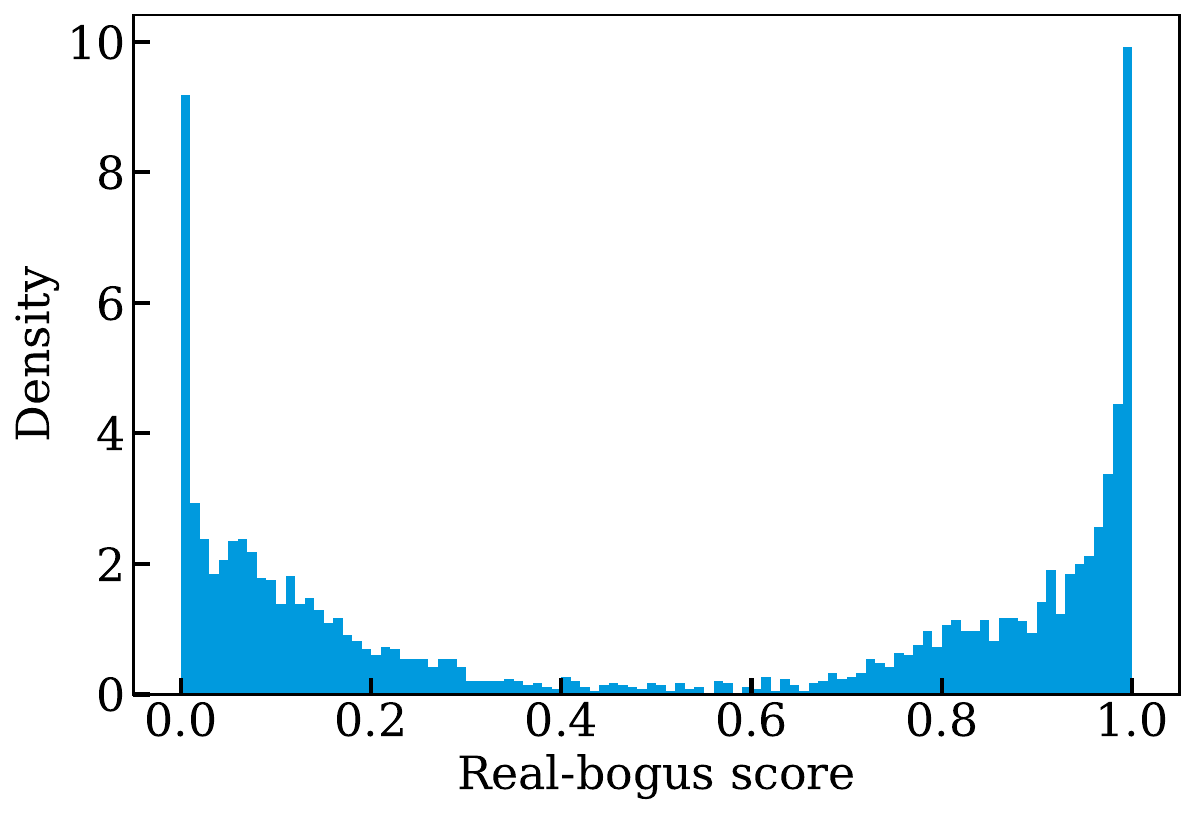}
\end{minipage}
\hfill
\begin{minipage}{0.495\linewidth}
\includegraphics[width=1\linewidth]{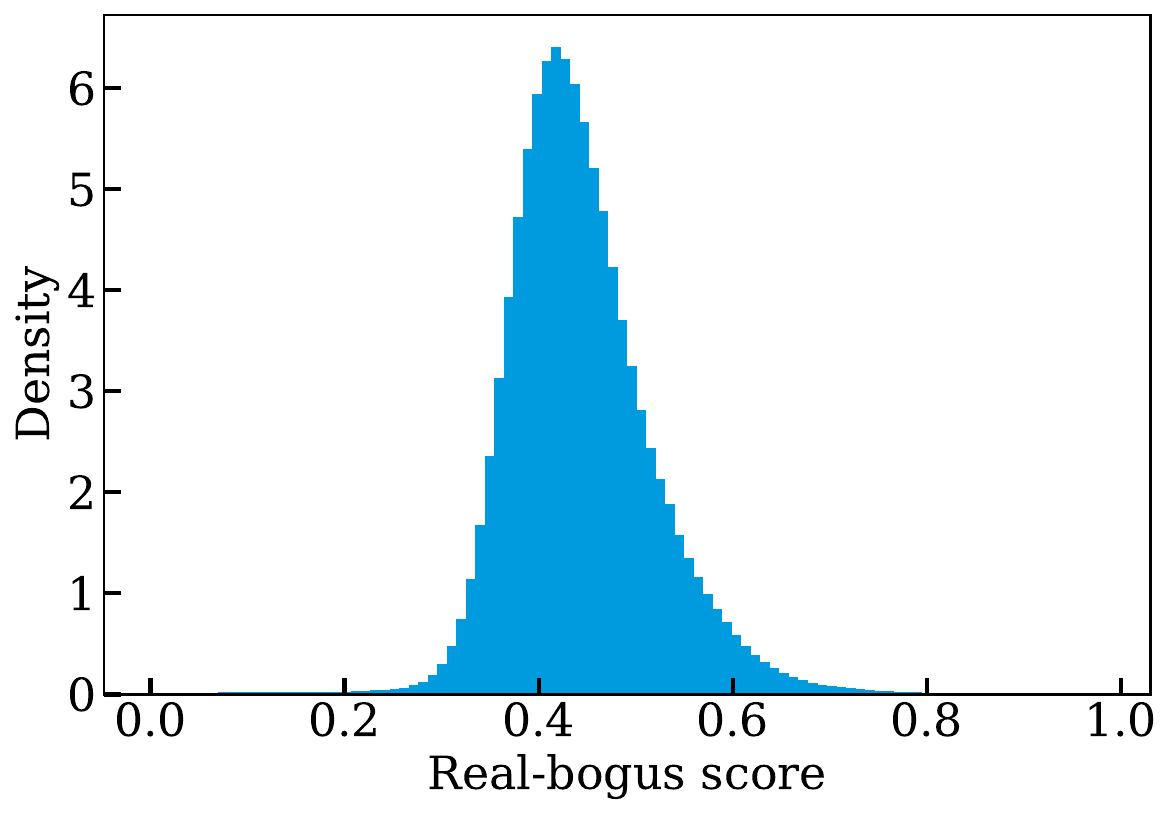}
\end{minipage}
\caption{Histograms of real-bogus classifier (Random Forest model) predictions. \textbf{Left:} objects from real-bogus dataset (3311 objects). \textbf{Right:} objects from target dataset (approximately 67 millions objects).}
\label{fig:rf_predictions}
\end{figure*}

Fig.~\ref{fig:rf_predictions} demonstrates  the  distributions of real-bogus score  based on the Random Forest model. The histogram of predictions for objects from the real-bogus dataset, as expected, has a bimodal distribution, while the distribution for target dataset resembles a bell curve. From this, we conclude that our classifier, trained on objects identified by SNAD experts as outliers in ZTF DRs, is capable of classifying such outliers. However, when the classifier see features of an object with a distribution not represented in the real-bogus dataset, it outputs a real-bogus score close to 0.5. This can be interpreted as the model being unsure about the class to which this object belongs. This reduces the likelihood of the classifier labeling an interesting anomaly as an artefact.

\section{Anomaly detection with real-bogus scores}
\label{sec:anomaly_detection}
In the SNAD pipeline, anomaly detection is carried out using algorithms based on random trees. The input data consist of light curve features, based on which the algorithm ranks objects according to their anomaly score. Then, a reasonable sample of the most anomalous objects is provided to the expert for manual verification. The sample size (budget) is a hyperparameter, which is defined by expert.

Integration of the real-bogus score into the anomaly detection algorithm is achieved by adding the classifier's prediction to the initial set of features (described in Section~\ref{subsec:lc}).  
We refer to the resulting set matrix as the \textit{augmented feature set}, i.e., with real-bogus score. Next, we feed this matrix to the anomaly detection algorithm.

Below we test two anomaly detection algorithms and feature sets:  Isolation Forest and Active Anomaly Detection, are applied on the initial feature set and on the augmented\footnote{\revx{The initial and augmented feature datasets are available in Zenodo, at \url{https://doi.org/10.5281/zenodo.14174666}.}}. Our goal is to understand how the addition of a new feature affects the anomaly detection \tim{process}.

\subsection{Isolation forest}
\label{subsec:if}
Isolation Forest (IF;~\citealt{isolationforest}) is an ensemble of random trees that allows for identifying anomalies within the data. The main assumption of this algorithm is that anomalies are isolated in the feature space from normal objects. The fewer random splits a tree needs to separate an object from the rest, the more anomalous it is considered. The anomaly score of the IF is inversely proportional to the average number of such splits across all trees in the ensemble. \revx{In this work, we use an IF with 100 trees, each one constructed using 256 objects randomly chosen objects and maximum depth of 8.}

\rev{We run the IF from the {\tt coniferest}\footnote{\url{https://github.com/snad-space/coniferest}} {\tt Python} library on both feature sets three times with different random states -- 42 for run №1, 53 for run №2, 100 for run №3. We manually verified 100 objects with the highest anomaly scores for both sets of features in each run. In all three cases for an augmented feature set the number of artifacts in the algorithm output is smaller, however, the difference increases with the overall number of artifacts in a corresponding run (see Table~\ref{runs}).}

\begin{table}
\caption{\label{runs} \rev{Summary of IF and AAD runs for three different runs. The number of artefacts found among 100 algorithm's outputs for the initial/augmented feature sets is shown.}}
\begin{center}
\begin{tabular}{|c|c|c|c|}
\hline
\diagbox{Algorithm}{Run} &  №1 & №2 & №3 \\
\hline
IF & 11/10 & 48/34 & 22/14\\
AAD & 27/3 & 59/21 & 12/15\\
\hline
\end{tabular}
\end{center}
\end{table}

\subsection{Active Anomaly Discovery}
Active Anomaly Discovery \citep{Das:16,das2017incorporating} is an algorithm based on the IF that adapts to expert actions on the fly. During the model initialization stage, an IF is trained. Then, the object with the highest anomaly score is sent for review to an expert who provides feedback on its classification (anomalous or nominal). The obtained feedback is used to update the base model by changing its parameters. Subsequently, the original data is fed into the modified IF. These steps are repeated until a certain number of reviewed objects (budget) is reached. This algorithm is flexible to changes in the expert requirements, as the user chooses what to consider as an anomaly and what to consider as a normal object (see example usage in \citealt{2021A&A...650A.195I,2023A&A...672A.111P}). 
\revx{We use AAD with 100 trees, each constructed using 256 randomly chosen objects and maximum depth of 8.}

As with IF, AAD\footnote{\url{https://github.com/snad-space/zwad}} was independently applied to the data represented by the initial feature set and the augmented \rev{with the same random states}. During the operation, the expert classified artefacts with the label ``NO'' (0), which corresponds to nominal samples, and with the label ``YES'' (1) -- to indicate real astrophysical events. For both feature sets, 100 objects \rev{in each run} were checked. The number of artefacts \rev{is given in Table ~\ref{runs}. It can be noticed that IF with an additional column does not reduce the percentage of artifacts as good as AAD (see Section~\ref{subsec:structure}).}

\begin{figure}
	\centering 
	\includegraphics[width=0.45\textwidth]{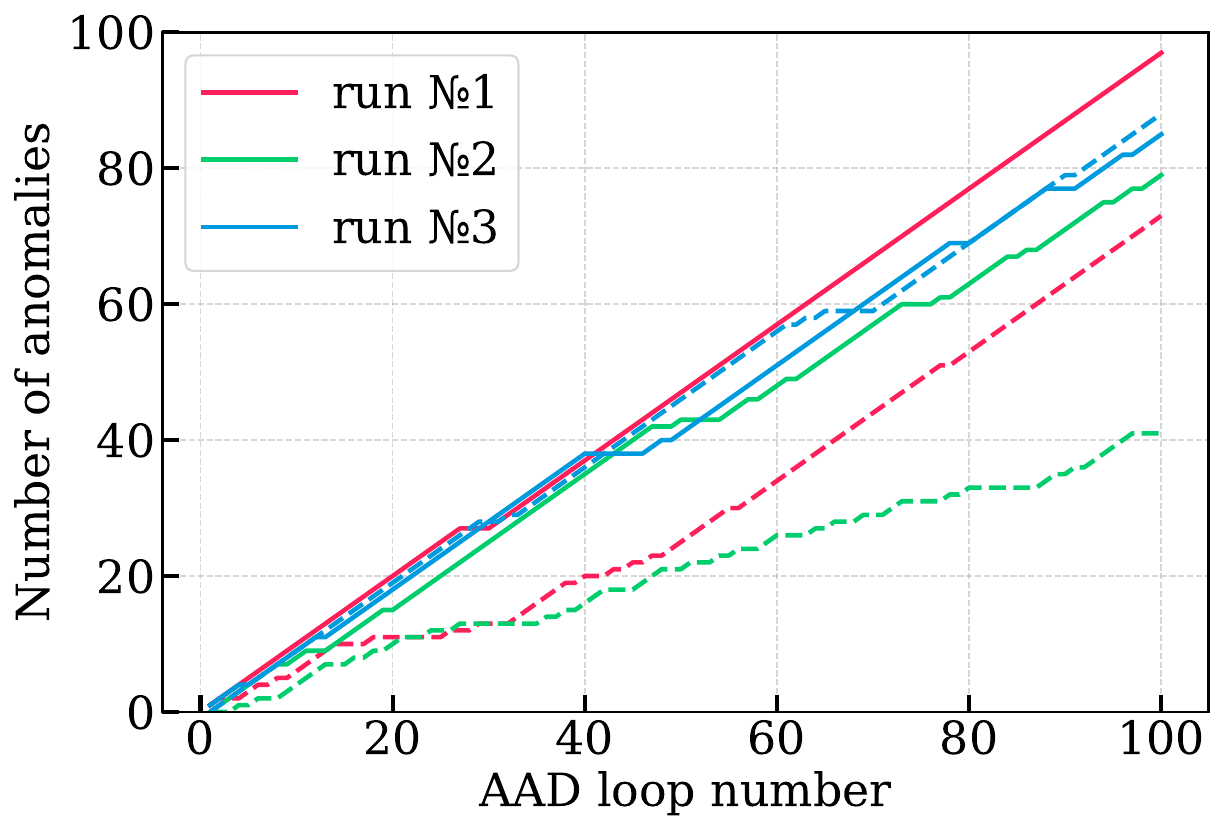}	
	\caption{Results of running AAD on two feature sets \rev{with different random states: number of anomalies versus number of verified candidates. Solid lines represent results for augmented feature set, while dashed lines stands for initial feature set.}} 
	\label{fig:aad}
\end{figure}

The dependency of the \rev{number} of anomalies on the number of candidates verified by the expert is presented in Fig.~\ref{fig:aad}. \rev{It can be seen that AAD, like IF, is sensitive to the choice of the random state. However, we observe a trend -- when the initialized isolation forest within the AAD focuses on a feature space region densely populated with artifacts, the addition of a new column significantly reduces the proportion of artifacts (run №1 and №2). Conversely, when it targets a region with fewer artifacts, then the number of artefacts differs by only a couple of percent, which has a negligible impact on the expert's time efficiency (run №3).}

\subsection{\tim{Forest structure}}
\label{subsec:structure}
\revx{The results presented in the previous section are a direct consequence of how the addition of one informative extra parameter impacts the construction of the forest, and how this change has  diverging consequences in the context of AAD and IF. AAD is concerned with the specific decision path within each tree in the calculation of anomaly scores, while IF only takes into account averages over the entire forest. Ultimately, this difference in describing the same decision tree structure will result in significantly different anomaly scores. Moreover, since AAD is an asymptomatic process, the cumulative effect throughout 100 loops becomes evident (Figure \ref{fig:aad}). }

This effect can be described by leaf vectors $z$~\citep{das2017incorporating}, \tim{which allow us to track} \revx{the particular leaf hosting each object in our dataset}.
For \revx{each object} $x$
there is a vector $z \in \{0, 1\}^{N}$, where $N$ is the number of leaves in
the IF ensemble. Each component of $z$ stands for particular leaf in IF,
$z_i=1$ means that instance $x$ fell into i-th leaf, while $z_i=0$ means the
opposite. By construction $\sum_{i=1}^{N}{z_i}$ equals to number of trees in
the IF.

\begin{figure*}
\centering
\begin{minipage}{0.495\linewidth}
\includegraphics[width=1\linewidth]{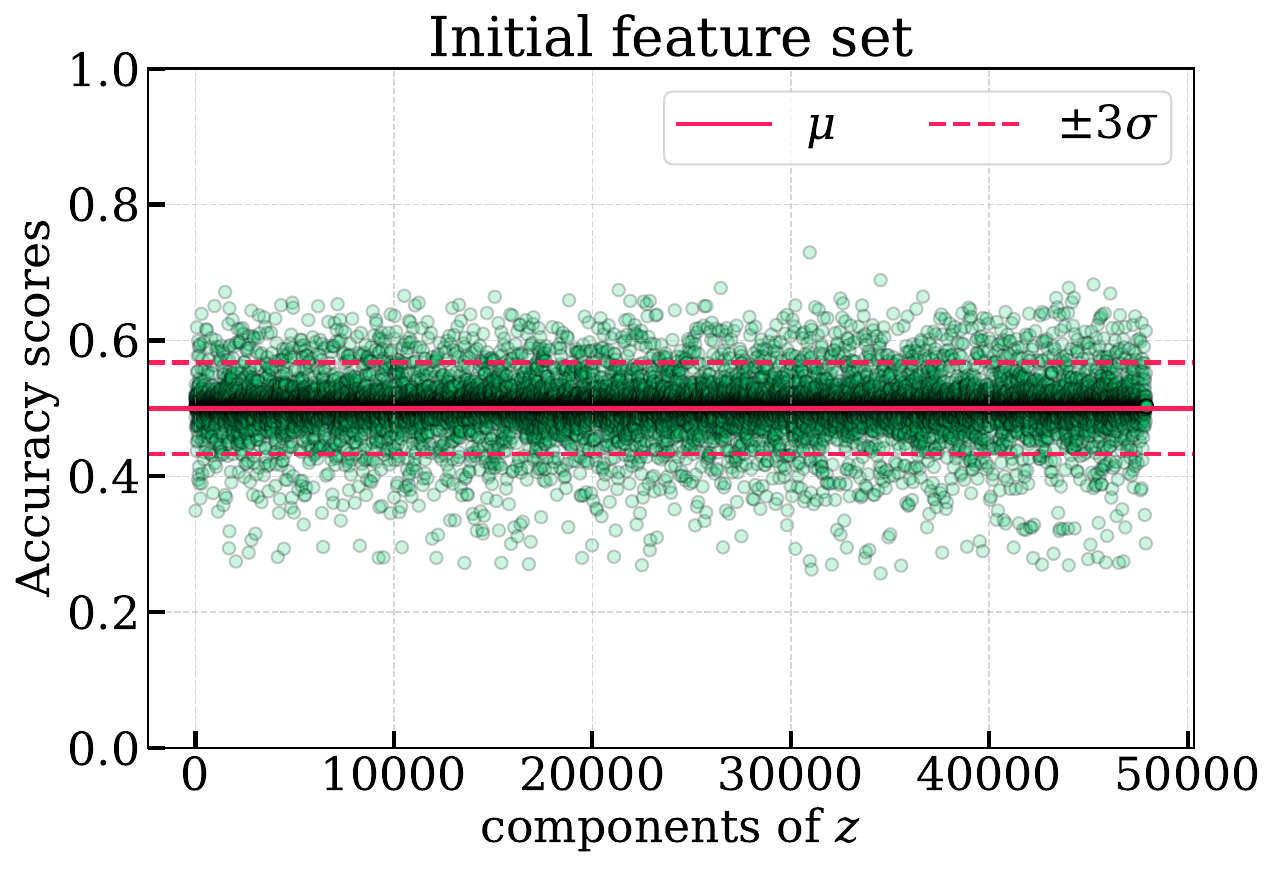}
\end{minipage}
\hfill
\begin{minipage}{0.495\linewidth}
\includegraphics[width=1\linewidth]{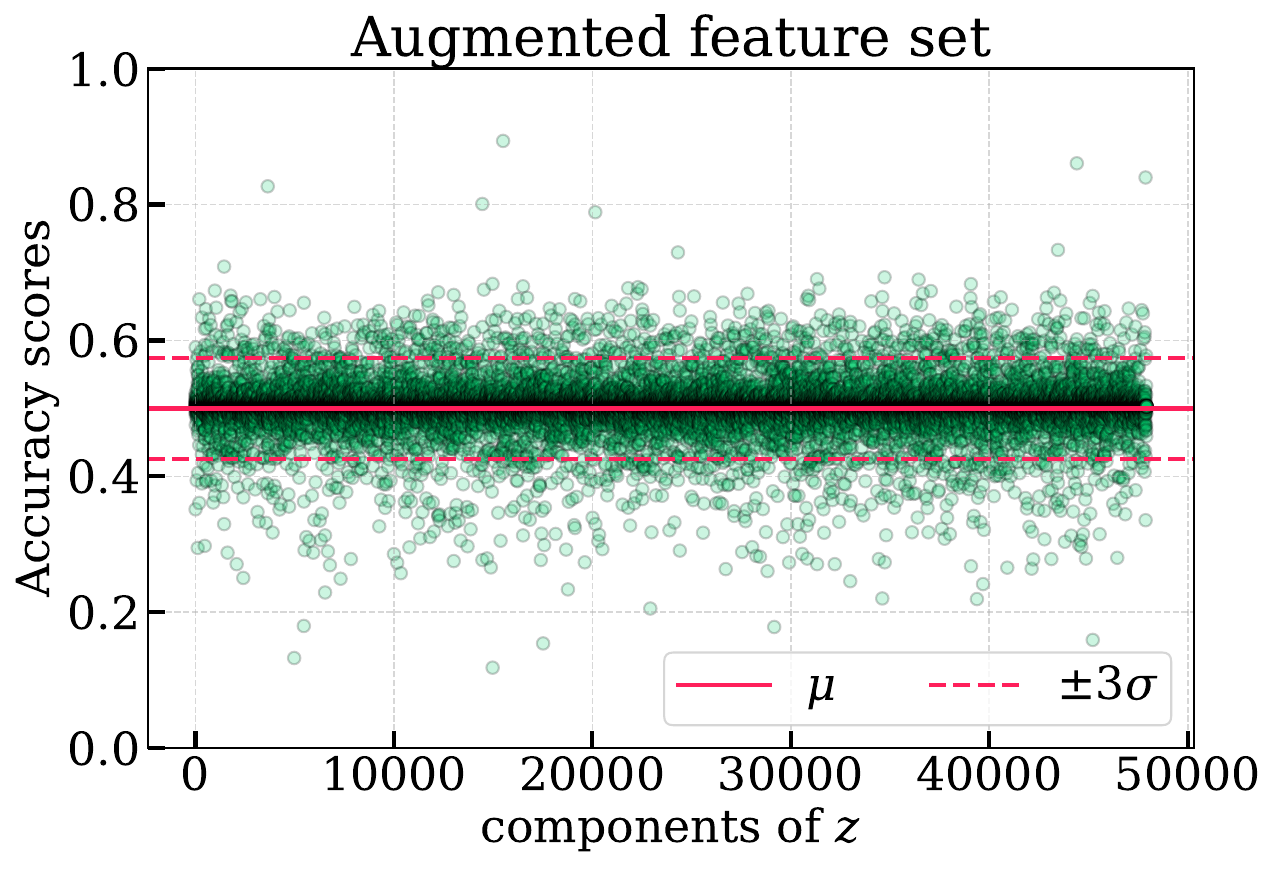}
\end{minipage}
\caption{Accuracy score for each leaf in IF \rev{(run №1)}. \textbf{Left:} Leafs from IF trained on initial feature set. \textbf{Right:} Leafs from IF trained on augmented feature set. Both IFs have 1000 trees. Although for the majority of leafs accuracy score stay the same when adding new feature, a few of them become significantly better discriminator -- larger number of extreme behavior in the right hand plot.}
\label{fig:zvectors}
\end{figure*}

We computed $z$ for each object from the real-bogus dataset represented by the
initial and augmented feature sets.
Then the confusion matrix can be estimated for every $z_i$ as if it was a real-bogus classifier prediction:
\begin{equation}
\mathrm{CM}_i = \begin{bmatrix}
\mathrm{TN}_i & \mathrm{FP}_i \\
\mathrm{FN}_i & \mathrm{TP}_i
\end{bmatrix},
\end{equation}
where
$\mathrm{TN}_i$ is the number of real objects with $z_i = 0$,
$\mathrm{FP}_i$ is the number of real objects with $z_i = 1$,
$\mathrm{FN}_i$ is the number of artifact objects with $z_i = 0$,
$\mathrm{TP}_i$ is the number of artifact objects with $z_i = 1$.
We use accuracy score:
\begin{equation}
\mathrm{Accuracy~score}_i = \frac{\mathrm{TN}_i + \mathrm{TP}_i}{M},
\end{equation}
(where $M$~-- number of objects in the real-bogus dataset) to explore artifact sensibility of every single leaf.
Essentially, this means that the higher the value of the accuracy score for a
leaf, the more often artefacts are encountered in that leaf and the less
frequently astrophysical objects are found there.
For the majority of leafs, the score is expected be $0.5$.
Fig.~\ref{fig:zvectors} shows the
accuracy scores for each leaf in the IF \rev{(run №1)}. According to this figure, it can be
observed that with additional information from the real-bogus classifier, the
IF occasionally creates leaves where artefacts are encountered much more frequently
(Average~score $> 0.8$). Conversely, the IF without classifier predictions does
not have such leaves as it was expected. 

The presence of such leaves in the AAD structure \rev{potentially affects the final result more significantly than in case of IF.} This is because the anomaly score in AAD is not calculated as the average number of splits needed to isolate an object, but as a weighted sum where the weights are the optimized parameters of the leaves. Therefore, AAD can effectively incorporate additional information from the classifier, which is confirmed in practice (Figure ~\ref{fig:aad}).

\section{\tim{Discussion}}
\label{sec:discussion}
\tim{While checking the outliers returned by AAD for each feature set, we found that only one real object was common between both sets (OID \texttt{601210400015694} -- \rev{a cataclysmic variable) in run №1, none in run №2, and one object (OID \texttt{708214400002778} -- a QSO) in run №3}. 
Also, three artifacts identified with the augmented feature set \rev{for the run №1 with a highest reduction in artefacts,} differ from those found with the initial feature set. These artifacts include a non-variable object's light curve distorted by a single bad photometric point caused by saturation (OID \texttt{414212400009210}), a galaxy affected by defocusing (OID \texttt{596212200016913}), and an artefact of the automatic ZTF photometry arising from incorrect background subtraction of the halo of a bright variable star (OID \texttt{646209400023624}; see also section 4.4.1 in~\citealt{2021MNRAS.502.5147M}).}

\tim{Therefore, we can conclude that incorporating additional information from the real-bogus classifier does indeed alter the distribution of  anomaly scores.} \revx{Nevertheless, in a few cases, }\tim{AAD could assign a low anomaly score to a real astrophysical event. As a result, using the described approach} \revx{does not guarantee that} \tim{potentially interesting objects \revx{will not be overlooked}. However, in the context of anomaly detection in big dataset like ZTF,} \revx{despite comprising only a very small fraction of the dataset, the absolute number of astrophysically interesting anomalies is not negligible.}
\tim{Although there is potential loss of some anomalies when using the augmented feature set, \rev{we achieved a reduction of artefacts in those cases when the feature space region is densely populated with artifacts.}}

\revx{Finally, we emphasize that the results presented here are bounded to the available computational and human resources currently available. We profit from the SNAD knowledge database, which was constructed through the efforts of a group of experts through half a decade of visual screening of ZTF data. Moreover, a large number of hours of expert time was required to search through both instances of the experiment described here. This screening is time consuming since it requires careful data mining and literature search to inform the algorithm about each candidate. In this context, the realization of multiple instances of the same experiment is not feasible. Nevertheless, the improvement in expert time allocation reported here, support by the theoretical description detailed in Section \ref{subsec:structure}, provide enough evidence to encourage further experiments within the same framework. Extended versions of this study including other classes and a more diverse set of extra features are under development and will be described in a subsequent work.}

\section{Conclusions}
\label{sec:conclusions}
In this work, we focus on reducing the fraction of artefacts presented to the expert by the AAD algorithm. 
We started from the SNAD knowledge database,  which consists of 3311 objects represented by light curve features \citep{2023PASP..135b4503M} and whose labels were assigned by experts after visual inspection. This data set was used  as a training set for 4 different real-bogus classifiers. Comparing their performance metrics, we selected the Random Forest model, for which the ROC-AUC~$= 0.94 \pm 0.01$. The classification scores obtained from this classifier were then used as an additional feature in the data input to two anomaly detection methods: static Isolation Forest and Active Anomaly Discovery.\rev{We found that additional information from the real-bogus classifier allows IF to reduce the number of artefacts in its outputs.}
\rev{Meanwhile, the IF does not reduce the percentage of artifacts as effectively as AAD. The rate at which AAD presents real objects to the expert is significantly accelerated when using a feature set that includes the real-bogus score, for random states corresponding to feature space regions densely populated with artifacts. For example, for random state 42 (run №1), the artifact ratio drops from 27 out of 100 with the initial feature set to 3 out of 100 with the augmented feature set. It is important to note that for some random states, using the augmented feature set does not reduce the number of artifacts and may even slightly increase it. In such cases, the total number of artifacts remains relatively low, and the inclusion of the additional feature does not significantly affect the time required for expert analysis of outliers.}
{\rev{However, the substantial reduction of artefacts in opposite cases} allows experts to focus their efforts on scrutinizing the astrophysical objects rather than artefacts, thereby greatly improving the efficiency of the review process}.

The approach proposed in this work can be applied to other classes of objects (e.~g., supernovae vs. non-supernovae). We can train several binary classifiers targeting  different classes and then add the resulting predictions as new features to the input data. The results shown in this work indicate that one can expect  significant speed up the rate to which AAD adapts to the feedback from the user.

The relevance of this work is highlighted by the forthcoming projects like LSST, which will generate an order of magnitude more data than ZTF. Filtering artefacts in such a data stream will become critically important.

\section*{Acknowledgements}
T.~Semenikhin, M.~Kornilov, A.~Lavrukhina, and A.~Volnova  acknowledges support from a Russian Science Foundation grant 24-22-00233, \url{https://rscf.ru/en/project/24-22-00233/} for conceptualization, software development  of the proposed algorithm; conducting experiments and analyzing of obtained results.
T.~Semenikhin acknowledges support  from the Theoretical Physics and Mathematics Advancement Foundation “BASIS” for  formalizing the proposed approach in a repository on GitHub and publishing the data on Zenodo.
Support was provided by Schmidt Sciences, LLC. for K. Malanchev.
E. E. O. Ishida received support for detailed review and editing of the article from  Universit\'e Clermont Auvergne through the grant AAP DRIF 2024 vague 1, ``Mobilités internationales sortantes''.
We thank the anonimnous referees for the provided feedback that helped to imrove the manuscript.

\bibliographystyle{elsarticle-harv} 
\bibliography{ref}

\end{document}